\documentstyle[12pt]{article}
\def\beq{\begin{equation}}
\def\eeq{\end{equation}}
\def\bea{\begin{eqnarray}}
\def\eea{\end{eqnarray}}

\newcommand\mpl{M_{\rm Pl}}

\catcode`\@=11
\def\marginnote#1{}
\def\ifmath#1{\relax\ifmmode #1\else $#1$\fi}

\def\U{{\cal U}}

\def\M{{\cal M}}
\def\wpsi{\widetilde{\Psi}}

\def\bold#1{\setbox0=\hbox{$#1$}%
     \kern-.025em\copy0\kern-\wd0
     \kern.05em\copy0\kern-\wd0
     \kern-.025em\raise.0433em\box0 }

\def\GENITEM#1;#2{\par\vskip6pt \hangafter=0 \hangindent=#1
   \Textindent{$ #2$ }\ignorespaces}

\newcount\hour
\newcount\minute
\newtoks\amorpm
\hour=\time\divide\hour by60
\minute=\time{\multiply\hour by60 \global\advance\minute by-
\hour}
\edef\standardtime{{\ifnum\hour<12 \global\amorpm={am}%
    \else\global\amorpm={pm}\advance\hour by-12 \fi
    \ifnum\hour=0 \hour=12 \fi
    \number\hour:\ifnum\minute<100\fi\number\minute\the\amorpm}}
\edef\militarytime{\number\hour:\ifnum\minute<100\fi\number\minute}
\def\draftlabel#1{{\@bsphack\if@filesw {\let\thepage\relax
  \xdef\@gtempa{\write\@auxout{\string
    \newlabel{#1}{{\@currentlabel}{\thepage}}}}}\@gtempa
    \if@nobreak \ifvmode\nobreak\fi\fi\fi\@esphack}
     \gdef\@eqnlabel{#1}}
\def\@eqnlabel{}
\def\@vacuum{}
\def\draftmarginnote#1{\marginpar{\raggedright\scriptsize\tt#1}}
\def\draft{\oddsidemargin -.5truein
        \def\@oddfoot{\sl preliminary draft \hfil
        \rm\thepage\hfil\sl\today\quad\militarytime}
        \let\@evenfoot\@oddfoot \overfullrule 3pt
        \let\label=\draftlabel
        \let\marginnote=\draftmarginnote

\def\@eqnnum{(\theequation)\rlap{\kern\marginparsep\tt\@eqnlabel}%
\global\let\@eqnlabel\@vacuum}  }
\def\preprint{\twocolumn\sloppy\flushbottom\parindent 1em
        \leftmargini 2em\leftmarginv .5em\leftmarginvi .5em
        \oddsidemargin -.5in    \evensidemargin -.5in
        \columnsep 15mm \footheight 0pt
        \textwidth 250mmin      \topmargin  -.4in
        \headheight 12pt \topskip .4in
        \textheight 175mm
        \footskip 0pt

\def\@oddhead{\thepage\hfil\addtocounter{page}{1}\thepage}
        \let\@evenhead\@oddhead \def\@oddfoot{} \def\@evenfoot{}
}
\def\titlepage{\@restonecolfalse\if@twocolumn\@restonecoltrue\o
necolumn
     \else \newpage \fi \thispagestyle{empty}\c@page\z@
        \def\thefootnote{\fnsymbol{footnote}} }
\def\endtitlepage{\if@restonecol\twocolumn \else  \fi
        \def\thefootnote{\arabic{footnote}}
        \setcounter{footnote}{0}}  
\catcode`@=12
\relax
\def\be{\begin{equation}}
\def\ee{\end{equation}}
\def\bea{\begin{eqnarray}}
\def\eea{\end{eqnarray}}
\def\simlt{\stackrel{<}{{}_\sim}}
\def\simgt{\stackrel{>}{{}_\sim}}

\def\mst11{m_{\;\widetilde{t}_{1}}}

\def\mst22{m_{\;\widetilde{t}_{2}}}
\def\mst12{m_{\;\widetilde{t}_{1,2}}}

\def\msb11{m_{\;\widetilde{b}_{1}}}
\def\msb22{m_{\;\widetilde{b}_{2}}}
\def\msb12{m_{\;\widetilde{b}_{1,2}}}

\def\mwidetilde2{\widetilde{m}^{2}}

\relax

\begin{document}
\input epsf

\topmargin-2.5cm
%
\begin{titlepage}
\begin{flushright}
astro--ph/0106022 \\
\end{flushright}
\vskip 0.1in
\begin{center}
{\Large\bf Oscillations During Inflation}

\vskip0.2cm

{\Large\bf and}

\vskip 0.2cm

{\Large\bf the Cosmological 
Density Perturbations}

\vskip .5in
{\large\bf N. Bartolo}$^{1,2}$,
{\large \bf S. Matarrese}$^{1,2}$ 
{\large and}
{\large\bf A. Riotto}$^{2}$

\vskip0.7cm
$^{1}${\it Dipartimento di Fisica di Padova ``G. Galilei''}

\vskip 0.2cm

{\it Via Marzolo 8, Padova I-35131, Italy}

\vskip 0.1cm
       
\begin{center}

{\it and}

\end{center}

\vskip 0.1cm

${^2}${\it INFN, Sezione di Padova}

\vskip 0.2cm

{\it Via Marzolo 8, Padova I-35131, Italy}

\end{center}
\vskip 1cm
\begin{center}
{\bf Abstract}
\end{center}
\begin{quote}

Adiabatic (curvature) perturbations are  produced during a period of
cosmological  inflation that is driven by a single  scalar field, the
inflaton.  On particle physics grounds -- though -- it is natural to expect
that this scalar field is coupled to other scalar degrees of freedom.
This gives rise  to oscillations 
between the perturbation of the inflaton field and the perturbations of
the other scalar degrees of freedom, similar to the phenomenon  of neutrino
oscillations. Since  the degree of the mixing 
is governed by the squared mass matrix of the scalar fields,
the oscillations  can occur 
even if the energy density of the extra  scalar fields is much smaller than
the energy density of the inflaton field. 
The probability of oscillation is resonantly
amplified when perturbations cross the horizon and the
perturbations in the inflaton field may disappear at horizon crossing
giving rise to  perturbations
in scalar fields other than the inflaton. Adiabatic 
and isocurvature  perturbations are inevitably
      correlated at the end of inflation and we provide a simple
expression  for the cross-correlation in terms of the slow-roll
parameters.

\end{quote}
\vskip1.cm
\begin{flushleft}
June 2001 \\
\end{flushleft}

\end{titlepage}
\setcounter{footnote}{0}
\setcounter{page}{0}
\newpage
%
\baselineskip=24pt
\noindent

\section{Introduction}

It is commonly   believed that the Universe underwent an
early era of cosmological inflation. The flatness and the  
horizon
problems of the standard big bang
cosmology are elegantly solved if, during the evolution of
the early
Universe, the energy density is dominated by some  vacuum energy
and comoving scales grow
quasi-exponentially. The  prediction of the simplest models of
inflation
is a flat Universe, {\it i.e.} $\Omega_{tot}=1$ with great precision.

Inflation \cite{guth81} has also become the dominant 
paradigm for understanding the 
initial conditions for structure formation and for Cosmic
Microwave Background (CMB) anisotropy generation. In the
inflationary picture, primordial density and gravity-wave fluctuations are
created from quantum fluctuations ``redshifted'' out of the horizon,
 where they
are ``frozen'' as perturbations in the background
metric \cite{muk81,hawking82,starobinsky82,guth82,bardeen83}. Metric 
perturbations at
the surface of last scattering are observable as temperature anisotropy in the
CMB.\\
Primordial perturbations can be of two kinds, the adiabatic and the isocurvature ones. Recently a lot of attention has been drawn on \emph{correlated} mixtures of the two \cite{langlois,LanRiaz}. In particular in Ref. 
\cite{LanRiaz} it has been shown how the correlation between the adiabatic and the isocurvature mode gives rise to new features both in the CMB anisotropies and in the large scale structure. These scenarios are strongly different from those usually consid
ered 
up to now, in which  only independent mixtures of the two modes were considered.\\      
Analyses of the possible constraints on these perturbations coming from the present data \cite{Trottaetal} and future experiments \cite{BucherMoodleyTurok} have been also made, as well as investigations on the production of the correlation during inflatio
n: 
in  Ref. \cite{langlois} a specific model of double inflation was considered, and in Ref. \cite{gw} a transparent formalism for studying the adiabatic and isocurvature modes was introduced.  
Here we pursue further the investigation on the possible correlation mechanism, trying to be as general as possible.\\
Our starting point is the simple observation that, if the inflaton field
couples to other scalar degrees of freedom,  oscillations 
between the perturbation of the inflaton field and the perturbations of
the other scalar degrees of freedom are induced.
For this phenomenon to happen, it is sufficient that the mass
squared matrix of the  scalar degrees of freedom is not
diagonal.  This induces a mixing among the
different scalar states and such a mixing  can be large even if the
energy density of the inflaton field dominates over the energy density
of the other scalars. We will show that the
 probability of oscillation is resonantly
amplified when perturbations cross the horizon. Adiabatic 
and isocurvature  perturbations are therefore
      correlated at the end of inflation and we provide a simple
expression  for the cross-correlation in terms of the slow-roll
parameters.

The plan of the paper is as follows. In Section $2$  we show that, during an inflationary period in which several scalar fields are present, is natural to expect a mixing and consequent oscillations between the fluctuations of the scalar fields. In Sectio
n 
$3$ the correlation of the adiabatic and isocurvature perturbations is explained in terms of this  
oscillation mechanism, and  an explicit expression for it is derived. Finally, in Section $4$, we show how to set the initial conditions  for structure formation  
in the post-inflationary epoch in the case where the isocurvature perturbations and the correlation are present.

\section{Oscillations during inflation}

Despite the simplicity of the inflationary paradigm, 
the number of inflation models
that have been proposed in the literature is enormous 
\cite{lrreview}. 
Models have been invented which predict  non-Gaussian density
fluctuations \cite{nongaussianinflation}, isocurvature fluctuation
modes \cite{isocurvatureinflation}, and cosmic
strings \cite{stringinflation}.

The simplest possibility is represented by the so-called 
 single-field models of inflation, where the 
the inflationary epoch can be  described by a single
dynamical order parameter, the  inflaton field $\phi$. 
Quantum fluctuations of the inflaton field produce Gaussian 
adiabatic perturbations of the  metric with a nearly scale
independent spectrum, $n_S\simeq 1$. The amplitude of the
perturbation can be characterized by the comoving curvature
perturbation ${\cal R}$, which remains constant on super-Hubble
scales until the perturbation comes back within the Hubble scale
long after inflation has ended. Single models
of inflation have already been  started to be constrained by the
recent accurate measurements of the CMB anisotropy \cite{kin}.

On particle physics grounds -- though --  it is hard  
to believe that  only one single scalar field $\phi$
plays a role  during the  inflationary stage. On the contrary, it
is quite natural to expect that during the inflationary dynamics 
several other scalar fields $\chi_I$ $(I=1,\cdots, N$) are present.
As 
soon as one considers more than one scalar field, one must also
consider the role of isocurvature fluctuations \cite{iso}. 
Such perturbations produce an anisotropy in the Cosmic Microwave
Background radiation  which is six times larger than the adiabatic
perturbations. To obtain small CMB anisotropy and still explain galaxy 
formation one has to strongly suppress isocurvature perturbations on the
horizon scale while keeping them sufficiently large on  galactic scales.
This can be done if the additional scalar fields acquire a mass 
of the order of the Hubble rate $H$ during inflation.
Furthermore, the presence of more than one scalar field may not only 
affect the evolution of the curvature perturbation, 
but also give rise to  the possibility of seeding isocurvature
perturbations after inflation. 

The contribution to the total energy density of the extra scalar
degrees of freedom $\chi_I$ might or might
not be negligible compared to the one provided by the 
scalar field $\phi$. If it is, the model of inflation is
called multiple field model, 
a general formalism to evaluate the
curvature perturbation at the end of inflation in such
models was developed in Ref.~\cite{SS}. 

However, even if  the contribution
to the total energy density of the scalar fields $\chi_I$
is small, quantum fluctuations $\delta\chi_I$ of the scalar fields
are amplified by gravitational effects. To compute the amplitude and the 
spectral shape of such perturbations one may consider
the theory of a single free scalar field  $\delta\chi_I$ with mass $m$
in a de Sitter background \cite{bd}.

In this section we wish to show that the generation of
the quantum fluctuations in the fields $\chi_I$ may be due to another
mechanism, which we call {\it oscillation mechanism} and  it is
important to stress that such a novel mechanism  
operates even if the energy density of the scalar field $\chi_I$ is
much smaller than the contribution coming from 
 a single scalar field $\phi$ which --
for such a reason --
deserves the name of inflaton field. 

If the scalar sector is composed by more than one single scalar
field $\phi$, all the  scalar degrees of
freedom will in general mix. In the particle physics language
this can be translated by saying that the mass squared (or the
Hamiltonian)
in the basis $(\phi,\chi_I)$ is not diagonal or, equivalently, that
 the
states $(\phi,\chi_I)$ are interaction eigenstates, but not mass eigenstates.
If the Hamiltonian of two quantum states is not diagonal, 
the interactions eigenstates   $(\phi,\chi_I)$ oscillate 
during the time evolution of the system. A familiar example in particle
physics is represented by the oscillating system of kaons $K^0$ and 
$\bar{K}^0$.

The oscillation mechanism responsible for the
amplification of the quantum fluctuations of the fields $\chi_I$ is 
due 
to  the  oscillations  of the inflaton fluctuations 
into fluctuations of the scalar fields $\chi_I$,  in the presence of the
inflaton background field $\phi_0$. In other words, a perturbation in the
inflaton field $\phi$ may evolve (oscillate) into a perturbation
of another scalar degree of freedom $\chi$ with a calculable probability.

From a pure
quantum mechanical point of view, such fluctuations  are generated
as coherent states. Therefore the oscillation mechanism is
a coherent production of quantum fluctuations\footnote{The
importance of coherent
production of particles after inflation during the preheating stage has
been recently emphasized in Ref. \cite{igor}.}. These oscillations
during inflation show a behaviour similar to the one
present  in the phenomenon of   coherent neutrino 
oscillations in a medium
like the Sun or the Earth. As we shall see, oscillations during inflation
are charecterized by amplification effects analogous to the MSW effect
of solar neutrinos.

Before launching ourselves into the details, 
let us now give a couple of examples supporting the
fact that a large mixing between  different scalar fields is generically
expected during the inflationary stage.
In  supergravity and (super)string models 
there  exists a plethora of 
scalar  fields -- loosely called  moduli -- 
with 
gravitational-strength couplings to 
ordinary matter. The mass of these scalar fields is of the
order  of the gravitino mass $m_{3/2}\sim$ 100 GeV in the present
vacuum, but is of the order of the Hubble rate $H$ during inflation. 
In these theories coupling 
constants and masses  appearing in the Lagrangian have to be thought as
functions of the dimensionless ratio $\chi_I/M_{{\rm Pl}}$, where
$\chi_I$ ($I=1,\cdots,N$) denotes a 
generic modulus field and $M_{{\rm Pl}}$ is the Planck mass.
For instance, a
generic
coupling constant $\lambda$  is in fact a function of the
scalar moduli
\begin{equation}
\lambda(\chi_I)=\lambda\left(\frac{\langle\chi_I\rangle}{M_{{\rm Pl}}} 
\right)\left(1 +c\frac{\delta\chi_I}{M_{{\rm Pl}}}
+\cdots\right),
\end{equation}
where $c$ is a coefficient usually of order unity and
$\delta\chi_I=\chi_I-\langle\chi_I\rangle$.
 This expansion introduces a 
direct coupling between the scalar moduli  and the inflaton. 
The potential $V$
becomes a function of two (or more) fields, $V=V(\phi,\chi_I)$ and  the second
derivative $(\partial^2 V/\partial\phi\partial\chi_I)$ may be as large
as $H^2$ during inflation. If we set $\phi\equiv \chi_0$, 
all the elements of the   mass
squared matrix ${\cal M}^2_{ij}\equiv (\partial^2 
V/\partial\chi_i\partial\phi_j)$ $(j=0,\cdots,N)$ are of the form $c_{ij}\,H^2$
where $c_{ij}={\cal O}(1)$.
A considerable mixing between  the
inflaton field and the moduli fields is generated if the inflaton 
background field $\phi_0$
takes values as large as the Planck scale. Notice that this may happen
even if $c_{00}\ll 1$, as required by the flatness of the inflaton 
potential. Under these circumstances, the perturbation of the scalar
field $\phi$ may oscillate into a perturbation of the modulus field
which is generated as a coherent state.

Another
example may be provided by 
 theories in which gravity may propagate in extra-dimensions 
\cite{extrareview} where there appears a infinite tower of spin-0  
graviscalar Kaluza-Klein excitations and  the inflaton field
can mix to these particles
by coupling to the higher-dimensional Ricci scalar. 

Let us now describe  a simple, but illustrative example. 
Consider two scalar fields, $\phi$ and $\chi$. We will dub $\phi$
the inflaton field, even if this might be a misnomer as the two fields
might give a comparable contribution to the total energy density
of the Universe. 

The scalar field perturbations, with comoving wavenumber $k=2\pi
a/\lambda$ for a mode with physical wavelength $\lambda$, 
obey the perturbation equations

\begin{eqnarray}
\label{m}
\delta\ddot{\phi} + 3H\delta\dot{\phi}
 + \frac{k^2}{a^2} \delta\phi +  V_{\phi\phi}
\delta\phi+V_{\phi\chi}
\delta\chi&=&0  \nonumber\\
\delta\ddot{\chi} + 3H\delta\dot{\chi}
 + \frac{k^2}{a^2} \delta\chi +  V_{\chi\chi}
\delta\chi+V_{\chi\phi}
\delta\phi&=&0,
\end{eqnarray}
where we have indicated by $V_{\phi\phi}=(\partial^2 
V/\partial\phi\partial\phi)$ and similar notation for the other
derivatives. 

The squared mass matrix is given by
\beq
\M^2=\left(
\begin{array}{cc}
V_{\phi\phi} & V_{\phi\chi}\\
V_{\phi\chi} & V_{\chi\chi}
\end{array}\right).
\eeq
We now introduce a time-dependent $2\times 2$ 
unitary matrix ${\cal U}$ such that 
\beq
\U^\dagger \M^2 \U={\rm diag}\,(\omega_1^2,\omega_2^2)\equiv \omega^2.
\eeq
In the following we will assume that 
all  the entries of the squared mass matrix
$\M^2$ are  real, so that 
 the unitary matrix $\U$ reduces to an orthogonal matrix
\beq
\label{u}
\U=\left(\begin{array}{cc}
\cos\theta & -\sin\theta\\
\sin\theta & \cos\theta
\end{array}\right),
\eeq
where
\beq
\tan 2\theta=\frac{2\,V_{\chi\phi}}{V_{\phi\phi}-V_{\chi\chi}}
\eeq
and the mass eigenvalues are given by 
\beq
\omega_{1,2}^2=\frac{1}{2}\left[\left(V_{\phi\phi}+
V_{\chi\chi}\right)\pm\sqrt{\left(V_{\phi\phi}-V_{\chi\chi}\right)^2
+4\,V_{\chi\phi}^2}\right].
\eeq
Adopting the vectorial notation $\Psi=(\Psi_1,\Psi_2)^T=
\U^T(\phi,\chi)^T$, 
the equations of the scalar perturbations may be rewritten
as 
\beq
\label{fund}
\delta\ddot{\Psi} + \left(3H+2\,\U^T\dot{\U}\right)
\delta\dot{\Psi}
 + \left(\frac{k^2}{a^2} +  
\omega^2+3\,H\,\U^T\dot{\U}+\U^T\ddot{\U}\right)
\delta\Psi=0,
\eeq
where 
\beq
\U^T\dot{\U}=
\left(\begin{array}{cc}
0 & 1\\
-1 & 0\end{array}
\right)\,\dot{\theta}
\eeq
and
\beq
\U^T\ddot{\U}=
\left(\begin{array}{cc}
0 & 1\\
-1 & 0\end{array}
\right)\,\ddot{\theta}+
\left(\begin{array}{cc}
-1 & 0\\
0 & -1\end{array}
\right)\,\dot{\theta}^2.
\eeq
Notice that the matrix $\U$  diagonalizes the squared mass matrix at the
price of introducing further non-diagonal terms in the equation of motion.

To proceed further, we may now take advantage of the  slow-roll conditions
which are to  be attained during inflation \cite{lrreview}. If we consider the 
generic slow-roll parameters
\beq
\varepsilon_{ij}=\frac{1}{2}\frac{M^2 V_i V_j}{V^2}\,\,\,\,
{\rm and}\,\,\,\, \eta_{ij}=M^2\frac{V_{ij}}{V},
\eeq
where $M=\mpl/\sqrt{8\pi}$ is the reduced Planck mass,
a successfull period of inflation requires that $\left|\varepsilon_{ij},
 \eta_{ij}\right|\ll 1$, {\it i.e.} the potential has to be
flat enough for inflation to develop. 

Since time derivatives of the
slow-roll parameters are second-order in the slow-roll parameters
themselves, $\dot{\varepsilon},\dot{\eta}\sim {\cal O}(\varepsilon^2,
\eta^2)$, it is easy to convince oneself that -- if we keep
only the slow-roll paramaters at the first-order -- Eq. (\ref{fund}) gets
simplified to 
\beq
\label{fund1}
\delta\ddot{\Psi} + 3H
\delta\dot{\Psi}
 + \left(\frac{k^2}{a^2} +  
\omega^2+3\,H\,\U^T\dot{\U}\right)
\delta\Psi=0.
\eeq
Introducing the conformal time $d\tau=dt/a$, where $a$ is the scale
factor of the expanding Universe, and the rescaled fields $\delta
\widetilde{\Psi}_1=
a\delta\Psi_1$ and $\delta\widetilde{\Psi}_2=a \delta\Psi_2$, Eq. (\ref{fund1})
becomes
\begin{eqnarray}
\label{pure}
\delta\wpsi_1^{\prime\prime}+\left(k^2-\frac{a^{\prime\prime}}{a}+
\omega_1^2 a^2\right)\delta\wpsi_1 +3H\,\theta^\prime a\,\delta\wpsi_2&=&0
\nonumber\\
\delta\wpsi_2^{\prime\prime}+\left(k^2-\frac{a^{\prime\prime}}{a}+
\omega_2^2 a^2\right)\delta\wpsi_2 -3H\,\theta^\prime a\,\delta\wpsi_1&=&0
\end{eqnarray}

To illustrate the phenomenon of oscillations during inflation, we
make the assumption that the non-diagonal terms in Eq. (\ref{pure})
proportional to $3H\theta^\prime a=3H\dot{\theta}a^2$ are smaller
than the diagonal entries $\omega_{1,2}^2 a^2$. This hypothesis is
correct, for instance, if the squared masses are constant in time.
We adopt this simplification in order
to render the description of the phenomenon of oscillations during inflation
more transparent.

In fact, in the following section we will solve the problem exactly 
at the first-order in the slow-roll parameters and show that at this order
of approximation taking  $3H\theta^\prime a\ll \omega_{1,2}^2$ does not
change the final result for the probability.

Eq. (\ref{pure})
is solved by\footnote{To be consistent one should also expand  the factor
$\frac{a^{\prime\prime}}{a}$ up to the first order in the slow-roll
parameters, reflecting the fact that inflation does not generically occurs
with  a pure de Sitter dynamics. However, these corrections
do not change the final expression for the oscillation probability, see Eq.
(\ref{fin}), since they would alter the wave-functions of the
two mass eigenstates in an equal manner.}
\beq
\label{sol}
\delta\wpsi_i=\frac{1}{2}\sqrt{\pi}\,e^{i\left(\nu_i+\frac{1}{2}
\right)\frac{\pi}{2}}\,(-\tau)^{1/2}\, H_{\nu_i}^{(1)}(-k\tau), \,\,\,
i=1,2, 
\eeq
where the conformal time $\tau$ assumes negative values (the
beginning of inflation is at some  $|\tau|\gg 1$), $ H_{\nu}^{(1)}$
are the Hankel's functions of the first kind and $\nu_i^2=9/4-(\omega_i/H)^2$.
The normalization factor
in Eq. (\ref{sol}) is chosen such that $\delta\wpsi_i$ matches the
plane-wave $e^{-ik\tau}/\sqrt{2k}$ for subhorizon scales in the
far ultraviolet $k/aH\gg 1$. 

We are now in the position to ask what is the probability (as a function 
of time) that a scalar
perturbation in the ``inflaton'' field $\delta\phi$ becomes a 
scalar field perturbation in the scalar field $\delta\chi$. 
The answer to this question is readily given if we remember that 
$[\phi(\tau),\chi(\tau)]^T=\U [\Psi_1(\tau),\Psi_2(\tau)]^T$. This means
that -- at a given time $\tau$ -- the scalar perturbations
$\delta\phi$ and $\delta\chi$ are a linear combination
of the mass eigenstates scalar perturbations $\delta\Psi_1$ and
$\delta\Psi_2$
\beq
\delta\phi=\sum_{\ell=1,2}\U_{1\ell}\,\delta\Psi_\ell,\,\,\,\,
\delta\chi=\sum_{\ell=1,2}\U_{2\ell}\,\delta\Psi_\ell.
\eeq
The probability that a scalar perturbation $\delta\phi$ at the time
$\tau_0$ becomes a scalar perturbation $\delta\chi$ at the time
$\tau$ is therefore given in general by
\beq
\label{prob}
P\left[\delta\phi(\tau_0)\rightarrow\delta\chi(\tau)\right]=
\left|\sum_{\ell=1,2}\U^{*}_{1\ell}(\tau_0)\,\U_{2\ell}(\tau)\,
\frac{\delta\Psi^*_\ell(\tau_0)}{\left|
\delta\Psi^*_\ell(\tau_0)\right|
}\frac{\delta\Psi_\ell(\tau)}{\left|\delta\Psi_\ell(\tau)\right|}\right|^2.
\eeq
We take as initial condition $\tau_{0} \rightarrow - \infty\, ,$ and we follow the conversion probability at \linebreak time $\tau$.\\
If the unitary matrix $\U$ is real, Eq. (\ref{prob})
becomes
\beq
P\left[\delta\phi(\tau_0)\rightarrow\delta\chi(\tau)\right]=
\frac{1}{2}\,\sin^2 2\theta\,\left\{
1-\frac{{\rm Re}\left[\delta\Psi_1^*(\tau_0)\delta\Psi_1(\tau)
\delta\Psi_2(\tau_0)
\delta\Psi_2^*(\tau)\right]
}{\left|\delta\Psi_1^*(\tau_0)\delta\Psi_1(\tau)\delta\Psi_2(\tau_0)
\delta\Psi_2^*(\tau)
\right|}\right\}.
\eeq
Let us investigate how such probability changes as a function of the
physical wavelength $\lambda=2\pi (a/k)$. At subhorizon scales,
$k\simgt aH$, the functions $\delta\Psi_k$ tend to the common 
plane-wave solution $e^{-ik\tau}/{\sqrt{2k}}$ and therefore
\beq
P\left[\delta\phi(\tau_0)\rightarrow\delta\chi(\tau)\right]\simeq
0\,\,\,(k\gg aH).
\eeq
However, at superhorizon scales, $k\simlt aH$, the functions
$\delta\wpsi_\ell$ develops a phase dependence 
\beq
\delta\wpsi_\ell\simeq e^{i(\nu_\ell - \frac{1}{2}) \frac{\pi}{2}}\,2^{\nu_\ell-\frac{3}{2}}
\,\frac{\Gamma(\nu_\ell)}{\Gamma(3/2)}\,\frac{1}{\sqrt{2k}}\,
(-k\tau)^{\frac{1}{2}-\nu_\ell}
\eeq
and the conversion probability becomes

\begin{eqnarray}
\label{fin}
P\left[\delta\phi(\tau_0)\rightarrow\delta\chi(\tau)\right]&=&
\frac{1}{2}\,\sin^2 2\theta\,\left[1-\cos\left(\frac{\pi}{2}\Delta\nu\right)
\right]\nonumber\\
&=&\sin^2 2\theta\,\sin^2\left(\frac{\pi}{4}\Delta\nu\right)
\,\,\, (k\ll aH),
\end{eqnarray}

where $\Delta\nu=\nu_1-\nu_2$.

In the limit $\omega^2_{1,2}\ll H^2$, such a probability 
reduces to

\beq 
P\left[\delta\phi(\tau_0)\rightarrow\delta\chi(\tau)\right]\simeq 
\sin^2 2\theta\,\sin^2\left(\frac{\pi}{12}\frac{\Delta\omega^2}{H^2}
\right),
\eeq

with $\Delta\omega^2=\omega_1^2-\omega_2^2$.

Note that the same formula holds for the 
probability $P\left[\delta\chi(\tau_0)\rightarrow\delta\phi(\tau)\right]$. 
This is due to the fact that eq. (\ref{prob}) depends only on the mass 
eigenstates and it is symmetric in $\delta \wpsi_1$ and $\delta \wpsi_2$.

The expression (\ref{fin}) reminds   the well-known formula 
which describes the evolution
in time of the probability of oscillations between two neutrino
flavours \cite{bil}. In both cases the probability  
is identically zero 
if the two mass eigenstates have equal
masses (no oscillations are present in the degenerate case); there
is the same dependence (as $\sin^2 2\theta$) on the mixing angle $\theta$ and
the same functional dependence on the difference of the
squared masses. Differences are present, though. While the probability
of neutrino conversion is depending upon time for any value of the
neutrino energy, in our case
at superhorizon
scales  the conversion probability becomes  constant
in time. This does not come as a surprise since on superhorizon scales
the dynamics of the system is frozen. 

What is more interesting is the time evolution of the conversion probability.
Consider a scalar perturbation in the inflaton field  with 
a given physical wavelength 
$\lambda=2\pi (a/k)$ which gets stretched during inflation. As long as the
wavelength remains subhorizon, the scalar perturbation remains
a pure perturbation in the inflaton field. However, as soon
as the wavelength crosses the horizon, the perturbation in the inflaton field
may become (generate) a perturbation in the other scalar field
$\chi$ with a nonvanishing probability determined by Eq. (\ref{fin}). 
At horizon crossing there is an amplification mechanism of the
fluctuations in the field $\chi$ which is reminiscent of the MSW effect
operative for solar neutrinos. 

The phenomenon of resonant amplification is easily understood if 
one remembers that a given wavelength crosses the horizon when $k=aH$,
{\it i.e.} when $k^2=a^{\prime\prime}/a$ using the conformal time. As
long as
the wavelength is subhorizon, $k^2\gg a^{\prime\prime}/a$, the presence of
the mass terms in the equations of motion (\ref{pure}) is
completely negligible compared to the factor $(k^2-a^{\prime\prime}/a)$.
On the other hand, when
the wavelength crosses the
horizon the term $(k^2-a^{\prime\prime}/a)$ vanishes and the effect of the
mixing in the mass squared matrix is magnified, giving rise to the
resonant effect. Finally, when the wavelength is larger than the horizon,
$k^2\ll a^{\prime\prime}/a$, the term $(k^2-a^{\prime\prime}/a)$ starts
to dominate again over the mass terms and the oscillations get frozen.

We conclude that fluctuations  in the scalar field $\chi$ are generated
as coherent states at horizon crossing 
through the oscillation mechanism out of 
perturbations in the scalar field $\phi$.

A couple of  comments are in order here. First of all, we wish to stress
that the oscillation mechanism operates even if the
energy of the inflaton field $\phi$ is much larger than the energy stored in
 the other scalar field $\chi$. This is because what is crucial for
the oscillations to occur is the {\it relative} magnitude of the elements
of the mass squared matrice ${\cal M}^2$. Secondly, the magnitude of the
probability depends upon two quantities, $\sin^2 2\theta$ and 
$\Delta\omega^2/H^2$. Both can be readily expressed in terms
of the slow-roll parameters.
The first factor is not necessarily small, in fact it may
be even of order unity for maximal mixing. If expanded in 
 terms of the slow-roll
parameters, it is ${\cal O}(\eta^0,\epsilon^0)$.
The second term is naturally
smaller than unity and is linear in the
slow-roll parameters. This reflects the fact that during inflation
only perturbations in those scalar fields with masses smaller 
than the Hubble rate may be excited. However, $\Delta\omega^2/H^2$
is not necessarily much smaller than unity and the amplification of
the conversion probability at horizon crossing may be sizeable.

\section{Correlation between adiabatic and entropy perturbations during
inflation}

As we already mentioned in the previous section,
adiabatic (curvature) and entropy (isocurvature)
perturbations are produced during a period of cosmological inflation if
more than one scalar field is present at this epoch. In this section we
wish to provide  a simple expression for   
the  cross-correlation between adiabatic and entropy
perturbations inspired by the considerations developed in the previous
section.

A consistent study of the linear field fluctuations 
requires the knowledge of the  linear scalar perturbations of the
metric, corresponding to the line element
\begin{eqnarray}
ds^2 &=& - (1+2A)dt^2 + 2aB_{,i}dx^idt \nonumber\\ &+&
a^2\left[ (1-2\psi)\delta_{ij} + 2E_{,ij}\right] dx^idx^j.
\end{eqnarray}

The consistent equation for a
generic  scalar field perturbation $\delta\chi_I$ $(I=1,\cdots,N$) 
with comoving wavenumber $k=2\pi
a/\lambda$ for a mode with physical wavelength $\lambda$ reads
\begin{eqnarray}
&& \ddot{\delta\chi}_I + 3H\dot{\delta\chi}_I
 + \frac{k^2}{a^2} \delta\chi_I + \sum_J V_{\chi_I\chi_J}
\delta\chi_J
  \nonumber\\ &=&
-2V_{\chi_I}A + \dot\chi_I \left[ \dot{A} + 3\dot{\psi} +
\frac{k^2}{a^2} (a^2\dot{E}-aB) \right].
\end{eqnarray}

At this stage it is useful to introduce the gauge-invariant Sasaki-Mukhanov
variables \cite{var}
\begin{equation}
\label{q}
Q_{I} \equiv \delta\varphi_I + \frac{\dot\varphi_I}{H}\psi
\end{equation}
which, in the spatially flat gauge $\psi_{Q}=0$, \ obey the equations of motion

\begin{equation}
\ddot{Q}_I + 3 H \dot{Q}_I + \frac{k^2}{a^2} Q_I  
+{\cal M}^2_{IJ} Q_J=0,
\end{equation}
where
\beq
\label{mass}
{\cal M}^2_{IJ}= V_{\chi_I\chi_J} - \frac{1}{M^2 a^3}
\left( \frac{a^3}{H} \dot \chi_I \dot \chi_J
\right)^{\cdot}\simeq \frac{V}{M^2}\left(\eta_{IJ}-
\frac{2}{3}\epsilon_{IJ}\right),
\eeq
where the last expression has been obtained performing an expansion up to
the first order in the slow-roll parameters.
This equation is similar to Eq. (\ref{m}) and from what we have learned in
the previous section, oscillations among the different quantities
$Q_I$ are expected to take place. Notice also that the $Q_I$'s  oscillate
even if the part of the mass squared matrix 
${\cal M}^2_{IJ}$ proportional to 
$V_{\chi_I\chi_J}$ is diagonal. This is because non-diagonal entries are
always present because of the $\epsilon_{IJ}$-parameters.

Once the variables $Q_I$ have been defined, one can
define 
the comoving curvature perturbation \cite{Lukash,Lyth85} 
\beq
\label{def:calR} 
{\cal R}=  \sum_I \left(\frac{\dot\varphi_I}{\sum_{J=1}^{N} \dot\varphi_J^2}
\right) Q_{I} 
\eeq
and give a  dimensionless definition of the total entropy perturbation
(automatically gauge-invariant) 
\begin{equation}
\label{S}
S = H \left( {\delta p \over \dot{p}} - {\delta\rho \over
\dot\rho} \right).
\end{equation}
For $N$ scalar
fields the latter is given by 
\begin{equation}
\label{sf} 
S= \frac {
2\left( \dot{V} + 3H
\sum_{J=1}^{N} \dot\varphi_J^2 \right) \delta V 
 + 
2\dot{V}\sum_I \dot\varphi_I (
\dot{\delta\varphi}_I - \dot\varphi_I A )
} {3
\left(2\dot{V}+3H\sum_J \dot\varphi_J^2\right) \sum_{I=1}^{N}
\dot\varphi_I^2}.
\end{equation}

Let us now restrict ourselves to the case of two fields, $\phi$ and $\chi$.
Following the nice treatement of Ref. \cite{gw}, we 
can 
define two 
new adiabatic and entropy fields
by a rotation in field space. We define
the ``entropy field'' $s$ \cite{gw}  
\beq
\label{s}
\delta s = (\cos\beta) \delta\chi - (\sin\beta) \delta\phi,
\eeq
where 
\begin{equation}
  \label{eq:cos sin}
\cos\beta = \frac{\dot{\phi}}{\sqrt{\dot{\phi}^2 +
\dot{\chi}^2}}, \quad \sin\beta =
\frac{\dot{\chi}}{\sqrt{\dot{\phi}^2 + \dot{\chi}^2}}.
\end{equation}
Notice that $\delta s$ can be rewritten as
\beq
\label{ss}
\delta s = (\cos\beta)  Q_\chi - (\sin\beta) Q_\phi,
\eeq
From this definition it follows that $s=$constant along the
classical trajectory, and hence entropy perturbations are
automatically gauge-invariant~\cite{StewartWalker,gw}. 

The adiabatic part of the perturbation is associated to the
orthogonal combination
\beq
\label{qqq}
\delta Q_A= (\sin\beta)  Q_\chi + (\cos\beta) Q_\phi.
\eeq
Our goal is to give an expression of the cross-correlation between
the adiabatic and the entropy perturbations

\begin{equation}
\langle Q_A(k) \delta s^*(k') \rangle \equiv {2\pi^2\over k^3}\, {\cal
C}_{Q_A\delta s} \, \delta(k-k').
\end{equation}
To do so, we adopt the technique developed in the previous section.
As we have seen, though, introducing a unitary matrix $\U$ which
diagonalizes the mass squared matrix is not enough to 
diagonalize the full system, (see Eq. (\ref{pure})). This happens because,
in general, the fields are coupled together.
To proceed, we first define the comoving fields $\widetilde{Q}_\phi=
a Q_\phi$ and $\widetilde{Q}_\chi=a
Q_\chi$, then we 
introduce a basis for annihilation and creation operators
${a_i}$ and ${a_i^\dagger}$ $(i=1,2)$ and perform the decomposition
($\tau$ is the conformal time)
\begin{eqnarray}
\label{deco}
\left(
\begin{array}{c}
\widetilde{Q}_\phi\\
\widetilde{Q}_\chi\end{array}
\right)
&=&\U\,\int\,\frac{d^3k}{(2\pi)^{3/2}}\,\left[e^{i{\bf k}\cdot{\bf x}}\,
h(\tau)\left(
\begin{array}{c}
a_1(k)\\
a_2(k)\end{array}\right)\,+\,{\rm h.c.}\right],\nonumber\\
\left(
\begin{array}{c}
\Pi_{\widetilde{Q}_\phi}\\
\Pi_{\widetilde{Q}_\chi}\end{array}
\right)
&=&\U\,\int\,\frac{d^3k}{(2\pi)^{3/2}}\,\left[e^{i{\bf k}\cdot{\bf x}}\,
\widetilde{h}(\tau)\left(
\begin{array}{c}
a_1(k)\\
a_2(k)\end{array}\right)\,+\,{\rm h.c.}\right],
\end{eqnarray}
where $\Pi_{\widetilde{Q}_\phi}$ and 
$\Pi_{\widetilde{Q}_\chi}$ are the conjugate momenta
of $\widetilde{Q}_\phi$ and 
$\widetilde{Q}_\chi$ respectively, and $h$ and $\widetilde{h}$ are 
two $2\times 2$ matrices  satisfying the relation
\beq
\left[h\,\widetilde{h}^* -h^*\,\widetilde{h}^T\right]_{ij}=
i\,\delta_{ij},
\eeq
derived from the canonical quantization condition. The matrix 
 $\U$ is given in Eq. (\ref{u}). As it  diagonalizes the 
squared mass matrix (\ref{mass}) 
(with the identification $\chi_1=\phi$ and $\chi_2=\chi$), 
the mixing angle is given by  
\beq
\tan 2\theta=\frac{2\,{\cal M}^2_{\chi\phi}}{{\cal M}^2_{\phi\phi}-
{\cal M}^2_{\chi\chi}}
\eeq
and the mass eigenvalues are given by 
\beq
\label{eig}
\omega_{1,2}^2=\frac{1}{2}\left[\left({\cal M}^2_{\phi\phi}+
{\cal M}^2_{\chi\chi}\right)\pm\sqrt{\left(
{\cal M}^2_{\phi\phi}-{\cal M}^2_{\chi\chi}\right)^2
+4\,{\cal M}^2_{\chi\phi}}\right].
\eeq
It is not difficult to see that the matrix $h$ satisfies the following
differential equation
\beq
h^{\prime\prime} + 2\,\U\,\U^\prime\,h^\prime
+\left[k^2-\frac{a^{\prime\prime}}{a}+
\omega^2 a^2+ \left(
\U\,\U^\prime\right)^2 +(\U\,\U^\prime)^\prime
\right]\,h=0,\eeq
If we now expand up to 
 the first order of perturbation in the slow-roll
parameters, we obtain
\begin{eqnarray}
\label{sys}
h_{11}^{\prime\prime} + 2\,\theta^\prime\,h_{21}^\prime
+\left(k^2-\frac{a^{\prime\prime}}{a}+
\omega_1^2 a^2\right)h_{11}+H\,\theta^\prime a \,h_{21}&=&0,\nonumber\\
h_{21}^{\prime\prime} - 2\,\theta^\prime\,h_{11}^\prime
+\left(k^2-\frac{a^{\prime\prime}}{a}+
\omega_2^2 a^2\right)h_{21}-H\,\theta^\prime a \,h_{11}&=&0,
\end{eqnarray}
and similar equations for $h_{12}$ and $h_{22}$ with the substitution
$h_{11}\rightarrow h_{12}$ and $h_{21}\rightarrow h_{22}$.

Making use of the decomposition (\ref{deco}) and definitions (\ref{ss})
and (\ref{qqq}), we find that
\begin{eqnarray}
\label{cor}
a^2\langle Q_A(k) \delta s^*(k') \rangle &=&
\left(s_\beta c_\theta-c_\beta s_\theta\right)\left(
c_\beta c_\theta+s_\beta s_\theta\right)\left[
\left|h_{22}\right|^2-\left|h_{11}\right|^2+
\left|h_{21}\right|^2 -\left|h_{12}\right|^2\right]\nonumber\\
&+&\left(
c_\beta c_\theta+s_\beta s_\theta\right)^2\left[h_{11}h_{21}^*+
h_{12}h_{22}^*\right]\nonumber\\
&-&\left(s_\beta c_\theta-c_\beta s_\theta\right)^2
\left[h_{21}h_{11}^*+
h_{22}h_{12}^*\right],
\end{eqnarray}
where we have made use of the shorthand notation 
$c_{\beta(\theta)}=\cos\beta(\theta)$ and  
$s_{\beta(\theta)}=\sin\beta(\theta)$.
To check this expression, we can consider the following limiting
cases:

\vskip 0.5cm
{\it i) No squared mass matrix}: If the 
squared mass matrix (\ref{mass}) is identically
zero,
$Q_\phi$ and $Q_\chi$ are already orthogonal states and mass eigenstates
and their time evolution is identical.
This means that there is no correlation between them and one expects
$\langle Q_A(k) \delta s^*(k') \rangle\propto s_\beta c_\beta\left(
\langle Q_\phi(k) Q_\phi^*(k') \rangle-\langle Q_\chi(k) 
Q_\chi^*(k')\rangle\right)=0$. 
This is confirmed by Eq. (\ref{cor}) since in this case
$c_\theta=1$, $s_\theta=0$, $h_{11}=h_{22}$ and $h_{12}=h_{21}=0$.

\vskip 0.5cm
{\it ii) Two noninteracting fields}: Suppose the system is composed by two
fields $\phi$ and $\chi$ with potential $V=\frac{1}{2}m^2(\phi^2+\chi^2)$.
In such a case, $c_\beta=s_\beta=1/\sqrt{2}$ and the squared mass matrix
(\ref{mass}) as the following structure
\beq
{\cal M}^2\propto\left(\begin{array}{cc}
1 & 1\\
1 & 1\end{array}\right),
\eeq
coming from the terms proportional to the $\epsilon$-parameters. This means
that the mixing angle $\theta=\pi/4$ and $c_\theta=s_\theta=1/\sqrt{2}$.
The system is already diagonalized by the unitary matrix $\U$ and
there is no need for the  nondiagonal elements of the matrix $h$, 
$h_{12}=h_{21}=0$. In such a case, the expression (\ref{cor})
predicts that $\langle Q_A(k) \delta s^*(k') \rangle$ vanishes.
This confirms the findings of Ref. \cite{langlois}.

To compute the cross-correlation between the adiabatic and the isocurvature
modes in a more general way, 
we can expand the solutions of the system (\ref{sys})
in powers of the slow-roll parameters.

In the far ultraviolet $k\gg aH$
the squared mass matrix is negligible and $Q_\phi$ and $Q_\chi$
are mass eigenstates. Therefore, we set the physical initial
conditions by posing  the unitary matrix $\U$  proportional to the unity matrix
and $h_{ij}=\delta_{ij}\,e^{-ik\tau}/\sqrt{2k}$. A simple inspection of the system 
(\ref{sys}) tells us that the functions $h_{12}$ and 
$h_{21}$ are sourced by the functions $h_{22}$ and $-h_{11}$, respectively.
Since 
$\theta^\prime$ is already ${\cal O}(\eta,\epsilon)$  during
the time evolution of the system 
$h_{12}=-h_{21}={\cal O}(\eta,\epsilon)$.
Expanding Eq. (\ref{cor}) up to  first order
in the slow-roll parameters, we get
\begin{eqnarray}
\label{cor1}
a^2\langle Q_A(k) \delta s^*(k') \rangle &=&
\left(s_\beta c_\theta-c_\beta s_\theta\right)\left(
c_\beta c_\theta+s_\beta s_\theta\right)\left[
\left|h_{22}\right|^2-\left|h_{11}\right|^2\right]\nonumber\\
&+&\left[h_{11}h_{21}^* - h_{21}h_{11}^*\right],
\end{eqnarray}
where we have made use of the fact that $
\left|h_{22}\right|^2-\left|h_{11}\right|^2={\cal O}(\eta,\epsilon)$.

A  simple perturbative procedure shows that
\begin{equation}
\label{onetwo}
h_{21}(\tau)\simeq
\int^\tau\, d\xi \,\frac{f(\xi)f^*(\tau)-f^*(\xi)f(\tau)}{W(\xi)}\, 
\theta^\prime(\xi)\,\left[2\, f^\prime(\xi)+H\,a(\xi)\, f(\xi)\right],
\end{equation}
where 
\begin{eqnarray}
f(\tau)&=&\frac{1}{2}\sqrt{\pi}\,e^{i\pi}\,
(-\tau)^{1/2}\, H_{3/2}^{(1)}(-k\tau),\nonumber\\
W(\tau)&=& f\, f^{*\prime}-f^{*}\,f^\prime.
\end{eqnarray}
As long as the wavelength of the perturbations is subhorizon, 
$h_{11}=h_{22}\simeq e^{-ik\tau}/\sqrt{2k}$ and solving Eq. (\ref{onetwo})
gives $\left|h_{21}\right|\propto (H/k)\,\left|h_{11}\right|\ll 
\left|h_{11}\right|$.

At superhorizon scales $k\ll aH$, $h_{11},h_{22}$ are proportional
to the scale factor (up to terms linear in the slow-roll parameters).
Since $\theta^\prime=a\,\dot{\theta}=a\,H\,F(\epsilon,\eta)$, where
$F$ is a function linear in the slow-roll parameters,
the integral in Eq. (\ref{onetwo}) can be performed and gives
\beq
h_{21}=-3\,h_{11}\,F(\epsilon,\eta)\,{\rm ln}\left(\frac{k}{aH}\right).
\eeq

With this information at hand, 
let us first evaluate the 
probability that a the perturbation
$Q_\phi$ at a time $\tau_0$ becomes a
perturbation
$Q_\chi$ at a generic time $\tau$. Using the decomposition (\ref{deco})
and making use of the fact that
$h_{12}=-h_{21}={\cal O}(\eta,\epsilon)$ during the time evolution
of the system, the  probability may be  written as
\begin{eqnarray}
&&P \left[Q_\phi(\tau_0)\rightarrow Q_\chi(\tau)\right] =
\left|\langle Q_\chi(\tau)|Q_\phi(\tau_0)\rangle\right|^2\nonumber\\
&=&\left|
c_\theta\, s_\theta\left[\frac{h_{11}(\tau_0)\,h^*_{11}(\tau)}
{\left|h_{11}(\tau_0)\,h^*_{11}(\tau)\right|}
-\frac{h_{22}(\tau_0)\,h^*_{22}(\tau)
}{\left|h_{22}(\tau_0)\,h^*_{22}(\tau)\right|}
\right]\right|^2.
\end{eqnarray}
Such a probability is vanishing at subhorizon scales, but at superhorizon
scale
\begin{eqnarray}
P\left[Q_\phi(\tau_0)\rightarrow Q_\chi(\tau)\right]&=&
\frac{1}{2}\,\sin^2
2\theta\,\left[1-\cos\left(\frac{\pi}{2}\Delta\nu\right)
\right]\nonumber\\
&=&\sin^2 2\theta\,\sin^2\left(\frac{\pi}{4}\Delta\nu\right),
\end{eqnarray}
which reproduces Eq. (\ref{fin}).

Therefore, we expect that the cross-correlation is extremely
small
at subhorizon scales reflecting
the fact that
$Q_\phi$ and $Q_\chi$
are good mass eigenstates
Nevertheless, the cross-correlation does not vanish at superhorizon scales
and is given by
\beq
{\cal C}_{Q_A\delta s}=
\left(s_\beta c_\theta-c_\beta s_\theta\right)\left(
c_\beta c_\theta+s_\beta s_\theta\right)
\left(\frac{H_k}{2\pi}\right)^2\,
\left[1-\left(\frac{k}{aH}\right)^{-2\Delta\nu}\right],
\eeq
where $H_k=k/a$ and $\Delta\nu=\nu_2-\nu_1$ (which is
given by  $-1/3\, \Delta\omega^2/H^2=-1/3(\omega_2^2-\omega_1^2)/H^2$ if
$\omega_{1,2}^2\ll H^2$).
If we normalize the scale factor $a$ in such a way that
$a=1$ at the end of inflation and indicate by $N=-{\rm ln}\,a$ the number
of {\it e}-folds until  the end of inflation, we finally get
\beq
{\cal C}_{Q_A\delta s}=
\left(s_\beta c_\theta-c_\beta s_\theta\right)\left(
c_\beta c_\theta+s_\beta s_\theta\right)
\left(\frac{H_k}{2\pi}\right)^2\,
\left[1-\left(\frac{k}{H}\right)^{\frac{2}{3}\frac{\Delta\omega^2}{H^2}}\,
e^{\frac{2}{3}N\frac{\Delta\omega^2}{H^2}}\right].
\eeq
This result is remarkably simple and can be easily
expressed in terms of the slow-roll parameters using the expressions
\beq
\frac{\Delta\omega^2}{H^2}\simeq 3\, 
\sqrt{\left[\eta_{\phi\phi}-\eta_{\chi\chi}
-\frac{2}{3}\left(\epsilon_{\phi\phi}-\epsilon_{\chi\chi}\right)\right]^2
+\, 4\,\left(\eta_{\phi\chi}-\frac{2}{3}\epsilon_{\phi\chi}\right)^2},
\eeq
and
\beq
\tan 2\theta=\frac{ 2\,\left(\eta_{\phi\chi}-\frac{2}{3}\epsilon_{\phi\chi}
\right)
}{\eta_{\phi\phi}-\eta_{\chi\chi}
-\frac{2}{3}\left(\epsilon_{\phi\phi}-\epsilon_{\chi\chi}\right)}.
\eeq

The origin of the cross-correlation is
due to a rather transparent physical behaviour. 
At the inflationary epoch, the 
gauge invariant perturbations $Q_\phi$ and $Q_\chi$
are generated with different wavelengths stretched by the
superluminal expansion of the scale factor. Since the
squared mass matrix of $Q_\phi$ and $Q_\chi$ is not diagonal, 
oscillations between the two quantities are expected.
Till the 
wavelength remains subhorizon, 
$Q_\phi$ and $Q_\chi$ evolve independently and may be 
considered good mass eigenstates. However, 
as soon as the wavelength crosses the horizon,  
an amplification in the probability of oscillation
between $Q_\phi$ and $Q_\chi$ occurs:  a nonvanishing correlation between
$Q_\phi$ and $Q_\chi$ is created on superhorizon scales
because of the nondiagonal mass matrix
${\cal M}^2_{IJ}$.
Since the adiabatic and the isocorvature modes are a linear combination
of  $Q_\phi$ and $Q_\chi$, at horizon crossing a nonvanishing
correlation between the adiabatic and the isocurvature modes is left imprinted
in the spectrum.\\
\section{Initial conditions in the radiation era}

It must  be emphasized that the mechanism described above for the production of adiabatic plus isocuvarture perturbations and their cross-correlation is active \emph{during inflation}. A common feature of isocurvature perturbations is that they might not

survive after the end of inflation \cite{Starobinsky,Linde,Polarstarobinsky}. If during reheating all the scalar fields decay into the same species (photons, neutrinos, cold dark matter and baryons), the only remaining perturbations will be of adiabatic t

ype.\\
In fact, in order to have isocurvature perturbations deep in the radiation era it is necessary to have at least one non-zero isocurvature perturbation \cite{KodSasa}:
\beq
S_{\alpha \beta}\equiv \frac{\delta_{\alpha}}{1+w_{\alpha}}-\frac{\delta_{\beta}}{1+w_{\beta}}
\neq 0,
\eeq 
where $\delta_{\alpha}=\delta
\rho_{\alpha}/\rho_{\alpha}$\, , $w_{\alpha}=p_{\alpha}/\rho_{\alpha}$ 
(the ratio of the pressure to the energy density), and $\alpha$ and $\beta$ 
stand for any two components of the system (they can also be scalar fields). 
$S_{\alpha \beta}$ (a gauge-invariant quantity) measures the relative 
fluctuations between the different components.  Adiabatic 
perturbations are characterized by having $S_{\alpha \beta}=0$ for all of 
the components. \\
So in the following we will assume that the mixing 
between the scalar fields is negligible after inflation and that, for 
example, the scalar field $\phi$ decays into ``ordinary'' matter 
(photons, neutrinos and baryons) and the scalar field $\chi$ decays only 
into cold dark matter (or $\chi$ does not decay, like the axion, 
so that it consitutes the cold dark matter). In this case we can write:
\beq
\delta_{CDM}=S_{CDM-{\rm rest}}+\delta_{A}\, , \qquad     
\delta_{A}=\frac{3}{4}\delta_{\gamma}=\frac{3}{4}\delta_{\nu}=\delta_{b}
\eeq
where $\delta_{A}$ specifies the amplitude of the adiabatic 
mode of perturbations, and ``rest'' stands fo photons, neutrinos and baryons.\\ 
In fact  the initial conditions for the evolution of cosmological 
perturbations are set once $S_{\alpha\beta}$ and the gravitational 
potential $\Phi$ (in the longitudinal gauge) are given deep in the radiation 
era \cite{langlois}. These initial conditions
are necessary to explore the effects on the large scale structure of the universe and the CMB  anisotropies from correlated adiabatic/isocurvature perturbations \cite{langlois, LanRiaz}.\\ 
Here we want to show how to link the value of $S_{CDM-{\rm rest}}$ and $\Phi$ to the inflationary quantities $\delta s$ and $Q_{A}$.\\
For  the gravitational potential the main equation is:
\beq
{\mathcal{R}}=- \frac{H}{\dot{H}} \dot{\Phi}+\left( 1-\frac{H^2}{\dot{H}} \right)\Phi,
\eeq 
where ${\mathcal{R}}$ is the curvature perturbation (see \cite{gw}, and reference therein).\\
Now, making an expansion in the slow-roll parameters to lowest order, it can be shown that the term proportional to $\dot{\Phi}$ can be neglected for most of the inflationary period. During the radiation dominated epoch $-{H^2}/\dot{H}=1/2$, and by matchi

ng the two stages we can write:
\beq
{\mathcal{R}}_{{\rm rad}}=\frac{3}{2} \Phi,
\eeq
where ${\mathcal{R}}_{{\rm rad}}$ is the curvature perturbation at the end of inflation, and it is directly related to $Q_{A}$ through \cite{gw}:
\beq
{\mathcal{R}}=Q_{A}\frac{H}{\dot{A}} \quad , \qquad     \dot{A}=(\cos \beta) \dot{\phi}+(\sin \beta) \dot{\chi}\, .
\eeq
As far as $S_{CDM-{\rm rest}}$ is concerned, let us define the following quantity:
\beq
\delta_{\chi \phi} \equiv \frac{\delta \chi}{\dot{\chi}}-\frac{\delta \phi}{\dot{\phi}}\, .
\eeq
For the two scalar fields $\phi$ and $\chi$ the isocurvature perturbation $S_{\chi \phi}$ as defined in Eq. $(53)$ results to be  $S_{\chi \phi}=a^3 {(\delta_{\chi \phi}/a^3)}^{\cdot}$ \cite{HwangNoh}. \\
On the other hand:
\beq
\delta s =\frac{\dot{\chi}\ \dot{\phi}}{\sqrt{\dot{\chi}^2+\dot{\phi}^2}}\ \delta_{\chi \phi} \, .
\eeq
Then, to lowest order in the slow-roll parameters and taking into account only the growing isocurvature mode, one finds:
\beq
S_{\chi \phi}=-\frac{3}{\sqrt{2}\, M}\ \frac{\sqrt{\varepsilon_{\phi \phi}+\varepsilon_{\chi \chi}}}{\varepsilon_{\phi \chi}}\ \delta s \, .
\eeq 
To match to the radiation epoch we take $S_{CDM-{\rm rest}}=S_{\chi \phi}$ at the end of inflation.\\
Finally, we can give an expression for the correlation between $\Phi$ and $S_{CDM-{\rm rest}}$ which represents the correlation between the adiabatic and isocurvature modes:
\beq
\langle \Phi(k) S^{*}(k') \rangle=-\frac{1}{8\pi\, M^{2}\varepsilon_{\phi \chi}}\, \langle Q_{A}(k)\delta s^{*}(k') \rangle,
\eeq 
where the right-hand side of this equation is evaluated at the end of inflation, and  $M=\mpl/\sqrt{8\pi}$.\\    
In a separate paper \cite{BarMatRio} we calculate in detail the amplitudes and spectral indeces in terms of the slow-roll parameters for $\Phi$, $S_{\alpha \beta}$ and the correlation.\\
\section{Conclusions}
In this paper we have shown that  the correlation between adiabatic and entropy perturbations during inflation \cite{langlois,gw} can be explained through an oscillation mechanism between the perturbations of two or more scalar fields.\\
If the isocurvature perturbation mode survives after inflation, this correlation can leave peculiar imprints on the CMB anisotropies as an additional parameter for the initial conditions of structure formation. In the near future very accurate data on  CM
B 
anisotropies will be available from  the  MAP \cite{MAP} and Planck \cite{Planck} experiments. Thus it is worth investigating further these issues, since they can represent a valid alternative to the simplest inflationary models. The latter are based on 
a single slow-rolling scalar field and generally predict adiabatic, scale-free and nearly Gaussian perturbations.\\
In this respect the relevant point of this work is to stress how isocurvature perturbations and cross correlations come out in a very natural way \emph{even} within a single-field inflationary scenario, with an inflaton $\phi$ which gives the dominant con
tribution 
to the total energy density and other 
scalar fields whose energy densities may 
or may not be subdominant. Moreover, the 
same correlation can give rise to a 
transfer of possible non-Gaussian 
features from the isocurvature mode to the adiabatic one. 
Different inflationary models, in fact, have been proposed in the 
literature which  predict non-Gaussian isocurvature perturbations 
\cite{nongaussianinflation}. 
In the present case, however, even if the isocurvature component 
is suppressed, as the observations 
suggest \cite{Trottaetal}, this transfer  mechanism can be 
very efficient for producing non-Gaussian \emph{adiabatic} perturbations. 
We will investigate this possibility in a separate paper \cite{bmr}.


\end{document}